# DYNAMIC CONTROL OF A FLOW-RACK AUTOMATED STORAGE AND RETRIEVAL SYSTEM

## Khalid HACHEMI*,**, Hassane ALLA*

* Laboratoire d'Automatique de Grenoble
ENSIEG – BP 46 F-38402 St Martin d'Hères Cedex, France
{khalid.hachemi, hassane.alla}@lag.ensieg.inpg.fr
** IMSI, Université d'Oran and Laboratoire d'Automatique de Tlemcen, Algeria

Abstract: In this paper we propose a control scheme based on coloured Petri net (CPN) for a flow-rack automated storage and retrieval system. The AS/RS is modelled using Coloured Petri nets, the developed model has been used to capture and provide the rack state. We introduce in the control system an optimization module as a decision process which performs a real-time optimization working on a discrete events time scale. The objective is to find bin locations for the retrieval requests by minimizing the total number of retrieval cycles for a batch of requests and thereby increase the system throughput. By solving the optimization model, the proposed method gives according to customers request and the rack state, the best bin locations for retrieval, i.e. allowing at the same time to satisfy the customers request and carrying out the minimum of retrieval cycles.

## 1. INTRODUCTION

A typical Automated Storage and Retrieval System is composed of storage racks, storage/retrieval (S/R) machines, conveyors and pickup/delivery (P/D) stations. There are several types of AS/RS such as unit-load, mini-load and flow-rack systems. The unit-load AS/RS is considered as the generic form of the others AS/RS. In the unit-load AS/RS, the items are stored and retrieved directly on the face of the rack by the storage/retrieval (S/R) machine. The flow-rack system is composed of deep bins which can contain several products. Two machines are used; a storage machine witch load items and a retrieval machine for items picking. These machines take place on opposite sides of the rack (storage side and retrieval side), and then neither function interferes with the other. So the items are stored also in-depth, which increases considerably the retrieval time of the products inside the rack. This fundamental difference between the two AS/RS due to in-depth storage induces a great difficulty of study and control. According to literature dealing with AS/RS, throughput system is considered the most important performance.

To achieve this goal, researchers developed several approaches. For a unit-load AS/RS and using a continuous rack face approximation, Bozer and White (1984) developed analytical expressions of travel time for single command cycle (SC) and dual command cycle (DC). They used randomized storage policy under a variety of input/output point and S/R machine dwell point configurations. Other authors were interested in optimal dwell point positioning of S/R machine (Egbelu, 1991; Peters *et al.*, 1996; Chang, and Egbelu, 1997). However, in a recent paper the authors in (Meller and Mungwattana, 2005) conclude, via a simulation study, that the dwell-point strategy has an insignificant impact on the relative system response time when the system is highly utilised. Sari, et *al.* (2005) developed closed-form travel time expressions for flow-rack AS/RS. Two approaches are presented; continuous and discrete. The continuous approach is based on a continuous rack face approximation. These expressions are compared, via simulation, with exact models given by the discrete approach. The authors found that there is no significant difference between the results obtained from the two approaches and conclude that the expressions based on continuous approach are easier to calculate and more practical than the discrete approach, which require extensive computation time. Another way to maximize throughput was developed by researchers. They were interested in scheduling methods witch sequence retrieval requests processed by the S/R machine in order to reduce the travel time. Generally, the storage requests are processed in first-come-first-served (FCFS) manner since the sequence of items to store present a physical flow where it is difficult to modify the filing (in a conveyor for example), while retrieval request can be re-sequenced because they are just an information flow (list). Lee and Schaefer (1996) studied a problem of sequencing retrieval requests in a unit-load AS/RS. In order to reduce the total travel time of the storage/retrieval machine, they introduced a method that finds either optimal or near-optimal solution quickly for moderate size problems. Van den Berg and Gademan (1999) were interested in the sequencing of storage/retrieval requests by considering the block sequencing approach, in an AS/RS with dedicated storage. The objective was to find a route of minimal total travel time. The problem was equivalent to the Travelling Salesman Problem; they showed that the particular case of sequencing under a dedicated storage policy can be solved in polynomial time. In (Kim, *et al.*, 2003), the authors proposed new replenishment process logic, according to the customer requests. The role of the replenishment process is to prepare items for the next picking cycle. The inputs of their replenishment model are demand for the next picking cycle and current pick face (location, product, remaining amount), while the output is the pick face for the next picking cycle (totes that will be replenished and number of compartments for each product type). The objective was to minimize the set-up time.

AS/RS control based on Petri nets received some attention in the literature. Amato, *et al.* (2005) developed two algorithms of control for an AS/RS modelled by Coloured Timed Petri Nets. They integrate an optimizer system in the control architecture witch performs real-time optimization to improve the throughput system. Dotoli, *et al.* (2005) proposed a modular modelling for an AS/RS comprising rail guided vehicles and narrow aisle cranes. They used Coloured Timed Petri Nets and tested several control and management policies to show how the simulated model can help to improve the overall system performance.

In this paper, we develop a dynamic control system for the flow-rack AS/RS and we are interested in the reduction of the system response time. This allows improving the system throughput, which corresponds to the number of products delivered per time unit. Most of the work on AS/RS was based mainly on the unit-load system, where the products are stored only on the rack face. In this system, for each requested product, just one retrieval cycle is done. For a flow-rack AS/RS, a requested product can be stored in depth inside the rack, thus the retrieval machine can requires more than one retrieval cycle to reach it. Therefore, we are also attached to the reduction of the restoring operations which are very expensive in term of time consuming.

With this intention, we started by modelling the AS/RS using Coloured Petri nets (CPN). We propose then a control and scheduling scheme which exploits the CPN model, as a state observer. We will see that the studied system behaves like a FIFO queue. The decision rules are generated by an optimization module of the number of retrieval cycles.

The paper is organized as follows. The flow-rack system and its operation are described in Section 2. Section 3 exposes the CPN model and its coupling with the optimization module. In Section 4, the mathematical model of the retrieval problem and a case study are presented. Finally, Section 5 gives conclusions.

## 2. FLOW-RACK AUTOMATED STORAGE AND RETRIEVAL SYSTEM

Gravity flow-rack AS/RS is composed of a deep rack with several bins. Each bin can contain multiple items placed according to a FIFO rotation. When an item is removed from a bin, the next item automatically rolls to the front of the rack. The rack uses inclined shelves equipped with roll track to move products by gravity from the storage to the picking (retrieval) side of the system.

The products are simply loaded into the back of the system (storage side) by the storage machine (SM) and they flow to the front of the system. The products are picked on the retrieval side by the retrieval machine (RM).

The storage and retrieval machines can move simultaneously on two axes. These two axes form the plan "xy" parallel with the two sides of the rack. A pickup station is located on the storage side, where the storage machine takes the products to be stored. A drop-off station is located on the retrieval side, where the retrieval machine deposits the products for the delivery. A restoring conveyor inclined in opposite direction, connects the retrieval machine to storage machine. It makes it possible to transfer the products to be restored towards the pickup station. Figure 1, gives the structure of a gravity flow rack used for distribution.

In Figure 2, the detailed configuration of a gravity flow-rack AS/RS is shown. The rack is composed of several bins, and each bin has several locations in which the products can be stored. Each location has a storage capacity of one item (product). For example, for retrieving a required product which is in the fourth location of a bin, the retrieval machine needs initially to retrieve one by one, the three products which precede it, to send them to the restoring conveyor for being reintroduced in the rack. The required product will be deposited in the drop-off station. The rack contains several types of products stored according to a policy of storage. We consider a random storage; i.e. a product can be stored in any location of the rack and a same bin can contain different products types. Given customers requests (the demand), the problem consists in determining bin locations of products satisfying this demand and realizing the minimum number of retrieval cycles.

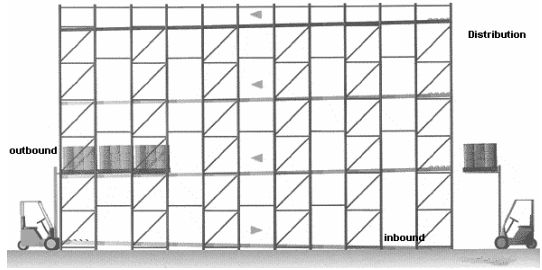

Fig. 1. Flow-rack structure

The request drawing up consists in gathering all customers' requests in only one batch represented by a vector. Each element of this vector represents the sum of item quantities of the corresponding product, resulting from all requests of customers.

For example for two customers requests $C_1$ and $C_2$ such as:

$C_1 = [q_{11} \quad q_{12} \dots\dots q_{1n}]^T$
$C_2 = [q_{21} \quad q_{22} \dots\dots q_{2n}]^T$

The resulting request vector $C$ will be equal to:

$C = C_1 + C_2 = [q_{11} + q_{21} \quad q_{12} + q_{22} \dots\dots q_{1n} + q_{2n}]^T$
$= [q_1 \quad q_2 \dots\dots q_n]^T$

With:

$q_{ij}$: amount of item j requested by customer i.

$q_j$: total amount of item j for all customers.

The advantage of gathering the requests in a batch is that the total number of retrieval cycles necessary to satisfy all the requests is shorter than that necessary if the processing is done request by request. That is due to the nature of the flow rack, where the successive retrieval operations are not independent. That can be shown easily for two requests for one item each one. The optimal solution necessarily does not correspond to the minimal index location of each item. Two operating modes can be considered according to the level of demand. In the first mode, the request drawing up is done at a fixed sampling period. In this case the bath size is constant. In the second mode, this period is adapted in order to reacts more quickly to the customers' requests, when the level of demand is important. In this case, the batch size is not constant.

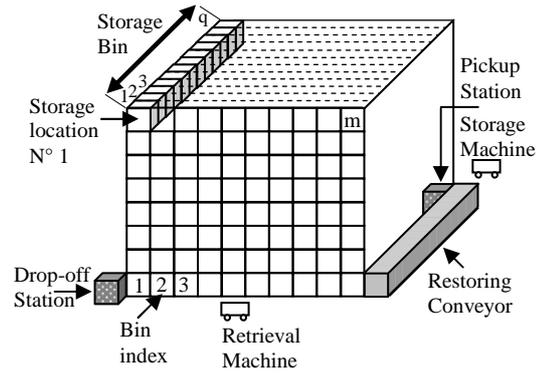

Fig. 2. Typical configuration for a flow-rack AS/RS

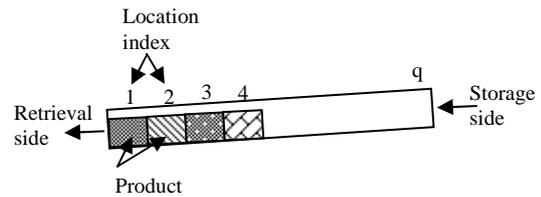

Fig. 3. Products layout inside the bin

# 3. STRUCTURE OF CONTROL SYSTEM

## 3.1 Coloured Petri net model

Coloured Petri nets (CPN) are derived from ordinary Petri nets and allow to model complex systems with identical sub parts. They constitute an abbreviation. In a CPN, with places are associated identifiers or "colours". Each transition can be fired in various ways represented by the various firing colours associated with the transition. Functions are associated with arcs; they define the relation between the firing colours and the coloured concerned marking. For more details, see (David and Alla, 2004). In order to represent the dynamics of the flow-rack AS/RS, a coloured Petri net is developed. Indeed, the system is of complex management, owing to the fact that the rack state at every moment can be modified by three different operations: the storage of a new product, the retrieval of a requested product and possibly the restoring of not requested products. Moreover, the size of the system is significant and it is necessary to take into account move of the products from a location to another inside the bins. The CPN model is shown in Figure 4. At any given moment, the system state is expressed by the CPN marking.

The model of Figure 4 represents the bins of the rack like FIFO queues. Indeed, the operation of a bin is similar to a FIFO, owing to the fact that the product which will be introduced first in the bin will leave it first, even if it is led once again to the restoring. The sets of colours which manage the CPN model are:

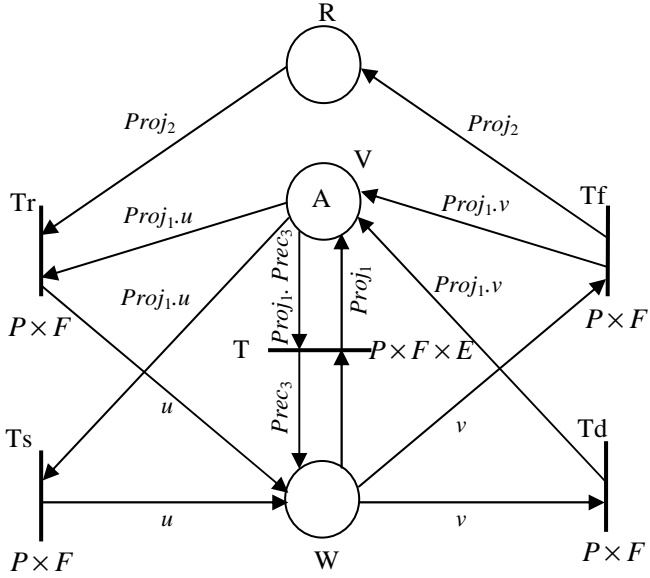

$$A = M_0(V) = \sum_{k=1}^{m} \sum_{j=1}^{q} < fk, ej >$$

Fig. 4. CPN model of flow-rack AS/RS

$P = \{ \prec pi \succ, i \in [1, n] \}$ : set of products type.

$F = \{ \prec fk \succ, k \in [1, m] \}$ : set of rack bins.

$E = \{ \prec ej \succ, j \in [1, q] \}$ : set of bin locations.

The arcs are associated functions. They are defined as follows:

The function $u$ introduces a product pi into the last location of a bin $fk$.

$u : P \times F \to P \times F \times \{eq\}$

$(pi, fk) \mapsto u(pi, fk) = \prec pi, fk, eq \succ$

The function v extracts a product pi from the first location of a bin $fk$.

$v : P \times F \to P \times F \times \{e1\}$

$(pi, fk) \mapsto v(pi, fk) = \prec pi, fk, e1 \succ$

Table 1 summarizes the used functions, related to the arcs (Alla, 1987)

### Table 1 Definition of the used functions

| Input set | Output set | Function | Definition |
|---|---|---|---|
| C | C | Id | Id(<ci>) = <ci> |
| C | <•> | Dec | Dec(<ci>) = <•> |
| $C_m$ | $C_m$ | $Prec_j$ | $Prec_j(<c_{i1}, c_{i2},..., c_{ij},..., c_{im}>)$ $= <c_{i1}, c_{i2},..., c_{(i-1)j},..., c_{im} > [n]$ |
| $C_m$ | $C_{m-1}$ | $Proj_j$ | $Proj_j(<c_{i1}, c_{i2},..., c_{ij},..., c_{im}>)$ $= <c_{i1},..., c_{i(j-1)}, c_{i(j1)},..., c_{im} >$ |

Note that C is a set of n colours: $C = \{ < c1 >, < c2 >,.., < Ci >,.., < Cn > \}$ and Cm is the Cartesian product with the order m of C.

The interpretation of places and transitions of the CPN model is summarized in Table 2.

| Places/Transition | Colours | Interpretation |
|---|---|---|
| V | <fk, ej> | The location $ej$ of the bin $fk$ is empty |
| W | <pi,fk, ej> | The location $ej$ of bin $fk$ contains a product of type $pi$ |
| R | <pi> | A product of type $pi$ is recycled |
| Ts | <pi, fk> | Introduce a product $pi$ in location $q$ of bin $fk$ |
| Tr | <pi, fk> | Introduce a recycled product $pi$ in location $q$ of bin $fk$ |
| Td | <pi, fk> | Evacuate a product of type $pi$ from location 1 of bin $fk$ towards the drop-off station |
| Tf | <pi, fk> | Evacuate a product of type $pi$ from location 1 of bin $fk$ towards the recycling |
| T | <pi,fk, ej> | Move a product $pi$ from location $j$ to location $j-1$ |

Table 2 Interpretation of CPN places and transitions

### 3.2 Feedback control

Figure 5 shows the AS/RS control based on CPN model. The reference input of the control system is the customers request periodically worked out. The values $q_1$ to $q_n$ represent the items (*1 to n*) quantities required after regrouping of the various customers requests.

The control system uses feedback from process sensors via the CPN model witch acts as state observer. That allows updating the state of the system after each operation creating a change of state, i.e. retrieval operation, storage operation and restoring operation. To achieve this goal, the CPN must be timed. The movement of the products from a location to another is very fast and it is of negligible duration compared to those of the storage and retrieval machines. Therefore, zero delays are associated with the CPN transitions.

Another feedback is eventually used to adapt the period of the request processing $T_s$ but will not be detailed here (shown in dotted lines in figure 5).

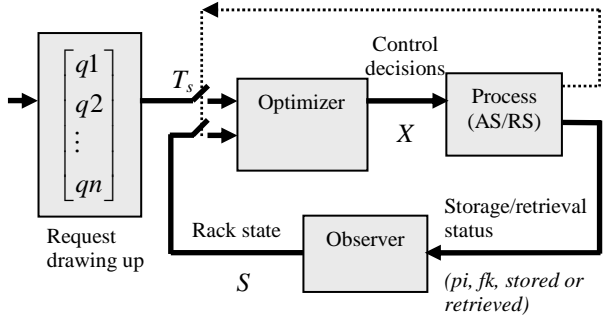

Fig. 5. Control scheme of the AS/RS

## 4. OPTIMIZATION BLOCK

The problem consists in satisfying the request of customers (treated by batch) by fulfilling the minimum of retrieval cycles. Indeed, a restoring operation of product is very expensive in consumption of time because it obliges to send back it to be reintroduced in the rack. Consequently, the difference between the access times to the various bins of the rack is not penalizing in terms of the number of retrieval cycles. So, we consider the entire penalty on the depth of the rack. Our objective is to minimize the number of cycles.

We consider a random storage; the products can be stored in any rack location. This choice allows to show the efficiency of the optimization method which however remains valid for any other storage policy.

The solution to search is to find the locations in the rack which contains the requested products and allowing the minimum of retrieval cycles. It is obvious that a minimum of retrieval cycles induces automatically a minimum of restoring cycles. Furthermore, the consideration of the number of retrieval cycles differs according to cases where the products are in the same bin, or in different bins.

To illustrate the retrieval problem, assume a rack with six bins of four locations each one. Figure 6 represents this rack on a plan x-y. This rack contains three types of products symbolized by ●, ■ and ▲. Assume a customers request, consisting of four products ● and two products ▲. A solution (trajectory) consists in determining a combination of bin locations which satisfies the batch request (Section 2). Two particular solutions are indicated in this figure (S1 indicated in full line and S2 in dotted lines).

To calculate the total retrieval cost, we calculate initially the cost by product and then we make the sum of costs for all the products. The cost of the product is equal to its depth in the bin. In the case of presence of several products of the trajectory in the same bin, the cost of all these products is equal to that of the deepest product. So, we will speak about the cost of the bin given by the deepest product. Therefore, the trajectory S1 has a total cost of: 1 + 2 + 3 + 3 + 4 + 4 = 17 while that of S2 is: 3 + 2 + 3 + 3 + 2 = 13 (only 5 terms appear).

Note that for the first solution, the number of products to be restored is equal to 17 - 6 = 11 products whereas only 13 - 6 = 7 products are to be restored for the second solution.

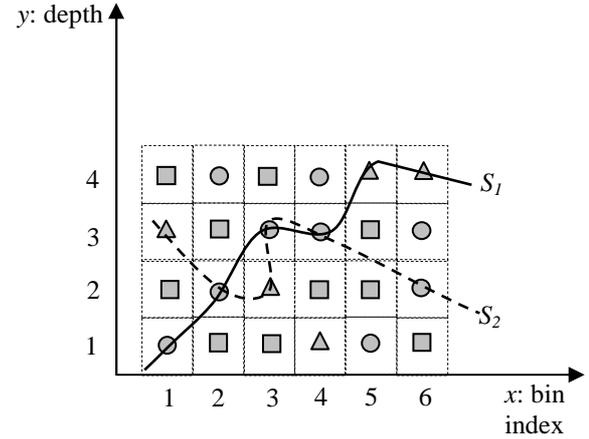

Fig. 6. Retrieval cost according to the selected locations

### 4.1 Mathematical formulation of the problem

Generally, the data are:

- $m$ bins with $q$ locations each one.
- Products of different types contained in the locations.
- The requested amounts of each product type.

**Problem:** Determine a trajectory satisfying a request batch (containing the requested products) of minimal total cost.

**Criteria:** Minimize the sum of the bins costs. This is formalized by Relations 1 - 8.

The rack is presented as a matrix where the lines indicate the indexes of bins and the columns the indexes of locations. The customers order (batch request) is represented by a vector which contains the quantities per type of required products. The problem thus consists in satisfying this demand with the minimum of retrieval cycles of the machine RM. The used parameters are:

$k$: bin index, $k \in [1, m]$.

$j$: location index, $j \in [1, q]$

$i$: product type index, $i \in [1, n]$

$S_{kj}$: rack state matrix. The elements of this matrix represent the products types stored in their respective locations. In fact, this matrix can be deduced from the marking M of the CPN model. This marking is updated according to the evolution of the system.

$$M = \begin{bmatrix} m(V) \\ m(W) \\ m(R) \\ m(D) \end{bmatrix}$$

$$S : F \times E \to P$$
$$(fk, ej) \mapsto S(fk, ej) = pi$$

*For each* $\prec pi, fk, ej \succ \in m(W): S(fk, ej) = pi$
$S(fk, ej) = \phi$ *otherwise*

By just using the indices:

$$S_{kj} = \begin{cases} i, & \text{if } \prec pi, fk, ej \succ \in m(W) \\ 0 & \text{otherwise} \end{cases}$$

$S_{kj} = 0$, it means that no product is in this location bin.

$C$: Vector of product type quantities requested by customers,

$M_{kj}$ : is a binary matrix definite as follows:

$M_{kj} = 0$ *or* $1$, $k = 1....m$ and $j = 1....q$

$X_{kj}$ : is an integer decision matrix definite as follows:

$X_{kj} = 0$ *or* $1$, $k = 1....m$ and $j = 1....q$

$X_{kj} = 1$, if the location j of bin k is selected for a retrieval

$X_{kj} = 0$, otherwise.

The objective can be formulated as follows:

$$Objective : \min \sum_{k=1}^{m} \sum_{j=1}^{q} M_{kj} * j \qquad (1)$$

**Subject to**

$$\sum_{j=1}^{q} M_{kj} \leq 1 \qquad \text{for each } k \qquad (2)$$

$$M_{kj} \leq 1 \text{ if } S_{kj} > 0 \qquad \text{for each } k, j \qquad (3)$$

$$M_{kj} = 0 \text{ if } S_{kj} = 0 \qquad \text{for each } k, j \qquad (4)$$

$$\sum_{k=1}^{m} \sum_{j=1}^{q} X_{kj} = C_i \text{ if } S_{kj} = i \quad \text{for each } i \qquad (5)$$

$$X_{kj} \leq 1 \qquad \text{if } S_{kj} > 0 \qquad \text{for each } k, j \qquad (6)$$

$$X_{kj} = 0 \qquad \text{if } S_{kj} = 0 \qquad \text{for each } k, j \qquad (7)$$

$$X_{kj} \leq \sum_{l=j}^{q} M_{kl} \qquad \text{for each } k, j \qquad (8)$$

The objective (1) consists in minimizing the retrieval cycles. It minimizes the total depth of products in the rack. Constraint (2) makes it possible to take into account the fact that, if several products are in the same bin, the depth expressing the number of cycles to be carried out for this bin is equal to the maximum between the products locations indexes in this bin. Constraints (3) an (4) allow to select the possible products locations to be retrieved. Constraint (5) enables to make a complete satisfaction of customers' request. Like constraints (3) and (4), Constraints (6) and (7) define the product locations to be selected but for the decision matrix X. Constraint (8) makes it possible to readjust the elements of matrix $M$, in order to allow the calculation of the number of retrieval cycles. Solving this model gives the products locations to be retrieved and the minimal number of retrieval cycles of the machine RM.

### 4.2 Case study

We consider a rack with six bins of seven locations each one. The rack contains 10 types of products, therefore:
m = 6; the number of rack bins,
q = 7; the number of locations in each bin,
n = 10; the number of the products type in the rack.
Since we are interested in optimization of the number of retrieval cycles and consequently the number of restoring cycles and for reasons of clarity, we will present the rack with bins arranged vertically. It is an abstract representation, equivalent to the physical representation.
Matrix S represents the rack state, such as $S_{kj} = i$, where $i$ is the product type in location $j$ of bin $k$. So, the elements of matrix $S$ represent the product types in each location of the respective bin. Figure 7 represents the rack according to the matrix $S$ considered in this example. The numbers inside the locations are the products types.

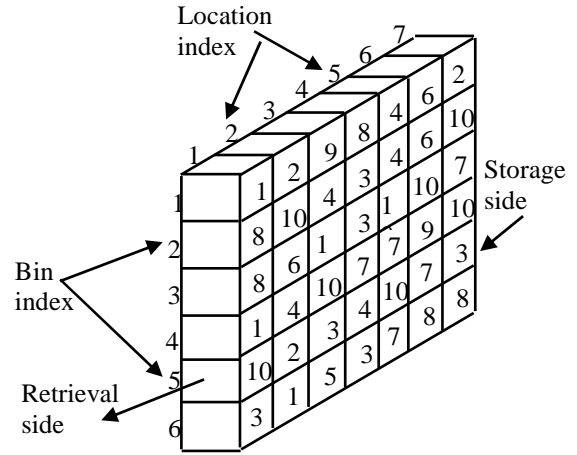

Fig. 7. Rack state with the types of products inside

Therefore,

$$S = \begin{bmatrix} 1 & 2 & 9 & 8 & 4 & 6 & 2 \\ 8 & 10 & 4 & 3 & 4 & 6 & 10 \\ 8 & 6 & 10 & 3 & 10 & 10 & 7 \\ 1 & 4 & 10 & 7 & 7 & 9 & 10 \\ 10 & 2 & 3 & 4 & 10 & 7 & 3 \\ 3 & 1 & 5 & 3 & 7 & 8 & 8 \end{bmatrix}$$

The customers' request is given in the form of a column vector, as follows:

$C = [3\ 3\ 0\ 5\ 0\ 0\ 0\ 0\ 0\ 5]^T$

The elements of this vector represent the quantities required for each type of product. This request consists of 3 products of type1, 3 of type 2, 5 of type 4 and 5 of type 10.

To create the software model of this optimization problem, we used AMPL language, which is a powerful, algebraic modelling language for problems in linear, nonlinear, and integer programming. The model is then solved by CPLEX; an optimization package for linear, network and integer programming.

After solving of the program using the AMPL/CPLEX system, we obtain the following results:

Objective = 24; it is the number of retrieval cycles to be carried out by the retrieval machine RM.

The decision matrix X:

$$X = \begin{bmatrix} 1 & 1 & 0 & 0 & 1 & 0 & 1 \\ 0 & 1 & 1 & 0 & 1 & 0 & 1 \\ 0 & 0 & 0 & 0 & 0 & 0 & 0 \\ 1 & 1 & 1 & 0 & 0 & 0 & 0 \\ 1 & 1 & 0 & 1 & 1 & 0 & 0 \\ 0 & 1 & 0 & 0 & 0 & 0 & 0 \end{bmatrix}$$

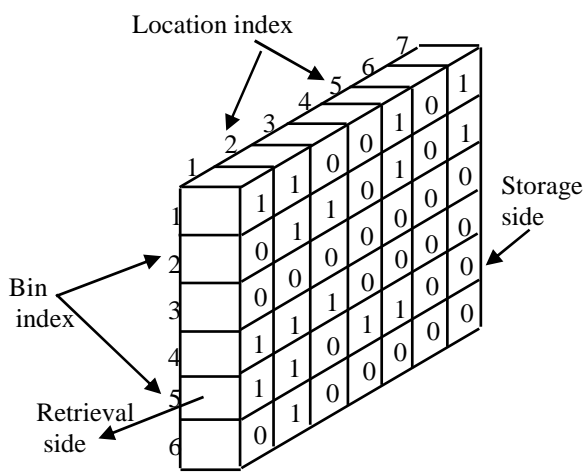

Fig. 8. Products locations for retrieval

The matrix $X$, in which the lines represent the bins indexes and the columns, the locations bins indexes, is a decision matrix with binary numbers; they can take either the value 0 or value 1:

0: The product contained in the location and the respective bin is not taken.

1: The product contained in the location and the respective bin is taken.

Figure 8 shows the products locations to be retrieved (locations containing value 1) according to the customers order. The control of retrieval machine is thus treated bin by bin according to table 3.

Table 3 Products locations for retrieval

| Bin ($fk$) | Products locations to be retrieved ($ej$) | Retrieval cycles number to be carried out (includes delivered and restored products) |
|---|---|---|
| F1 | e1, e2, e5, e7 | 7 |
| F2 | e2, e3, e5, e7 | 7 |
| F3 | No location | 0 |
| F4 | e1, e2, e3 | 3 |
| F5 | e1, e2, e4, e5 | 5 |
| F6 | E2 | 2 |

Note that since the retrieval machine RM, reaches only the first location of each bin, its control consists in making a certain number of retrieval cycles, for each bin. The number of retrieval cycles to carry out $c_k$ (includes the cycles of delivery and the cycles of restoring), for each rack k is equal to:

$$c_k = \sum_{j=1}^{q} M_{kj} * j$$

Table 4, gives the products locations to be retrieved by type. In this example, we see that the total number of cycles to carry out by the machine RM is equal to 24 cycles and the customers request is made up of sixteen products, that wants to say, that the number of products to be restored is equal to: 24 - 16 = 8 products, therefore to return to the restoring conveyor. The delivery rate is calculated as follows:

$$\tau_d = \frac{number\ of\ delivred\ products}{number\ of\ retrieved\ products} = \frac{16}{24} = 0.67$$

It means that 67% of retrieval operations are for delivery. The remainder represents the operations towards the restoring conveyor. Since the solution of the problem is optimal, this rate is the best delivery rate that it is possible to reach for the customers' request considered here.

Table 4 Products locations for retrieval by type

| Product Type | Requested quantities | Product locations to be retrieved (bin, location) |
|---|---|---|
| 1 | 3 | (f1, e1), (f4, e1), (f6, e2) |
| 2 | 3 | (f1, e2), (f1, e7), (f5, e2) |
| 4 | 5 | (f1, e6), (f2, e3), (f2, e5), (f4, e2), (f5, e4) |
| 10 | 5 | (f2, e2), (f2, e7), (f4, e3), (f5, e1), (f5, e5) |

## 5. CONCLUSIONS

In this paper, we have presented a control scheme of a gravity flow- rack AS/RS, based on a coloured Petri net (CPN). The proposed CPN model of the AS/RS, makes it possible to model dynamics of the system and thus to apprehend its complexity. In AS/RS, one of the most important performances is the system throughput. We were interested in minimizing the total number of retrieval cycles and consequently the number of restoring cycles which induce a very expensive processing time. With this intention we have integrated an optimization module as conflict solver, in order to select the products locations to be retrieved in the rack. The optimization problem is formulated in term of integer program. This module exploits the CPN marking to read the rack state. The solution of the problem gives the bin locations to be retrieved for a batch of customers' request and the minimal number of retrieval cycles of the retrieval machine. Note that the advantage of the given optimization method, is that it gives the exact minimal number of retrieval cycles. Thus, the delivery rate of the retrieval machine is maximized and its utilisation rate is minimized, which enables it to function more effectively with a higher system throughput rate. As perspective to this work, we intend to integrate the expiry dates of products in the sequencing problem of retrieval requests where in addition to the minimization of the number of retrieval cycles, it will be necessary to minimize also the risk of products expiry in the rack.